# Reuse of designs:
# Desperately seeking an interdisciplinary cognitive approach

Willemien Visser* & Brigitte Trousse°

INRIA (National Institute for Research in Computer Science and Control)

\* Unité de Rocquencourt
Ergonomic Psychology project
Rocquencourt B.P.105
78135 Le Chesnay Cedex (France)

tel: + 33 - 1 - 39 63 52 09
fax: + 33 - 1 - 39 63 53 30
email: visser@nuri.inria.fr

° Unité de Sophia-Antipolis
SECOIA project - Experts systems and Design of tools in AI
B.P. 93 - 2004, Route des Lucioles
06902 Sophia-Antipolis Cedex (France)

tel: + 33 - 93 - 65 77 45
fax: + 33 - 93 - 65 77 83
email: trousse@sophia.inria.fr

Abstract. This text analyses the papers accepted for the workshop "Reuse of designs: an interdisciplinary cognitive approach" (Visser, 1993). Several dimensions and questions considered as important (by the authors and/or by us) are addressed: What about the "interdisciplinary cognitive" character of the approaches adopted by the authors? Is design indeed a domain where the use of CBR is particularly suitable? Are there important distinctions between CBR and other approaches? Which types of knowledge -other than cases- is being, or might be, used in CBR systems? With respect to cases: are there different "types" of case and different types of case use? which formats are adopted for their representation? do cases have "components"? how are cases organised in the case memory? Concerning their retrieval: which types of index are used? on which types of relation is retrieval based? how does one retrieve only a selected number of cases, i.e., how does one retrieve only the "best" cases? which processes and strategies are used, by the system and by its user? Finally, some important aspects of CBR system development are shortly discussed: should CBR systems be assistance or autonomous systems? how can case knowledge be "acquired"? what about the empirical evaluation of CBR systems? The conclusion points out some lacking points: not much attention is paid to the user, and few papers have indeed adopted an interdisciplinary cognitive approach.

Interest of reusing "old" particular solutions (cases) in problem-solving tasks has, in recent years, led to proposals for systems based on case-based reasoning (CBR). That is why a workshop has been organised around the theme "Reuse of designs: an interdisciplinary cognitive approach" (Visser, 1993). This text introduces and analyses the papers presented at this workshop.

If "reuse" of designs has been studied -and implemented in (prototype) systems- particularly in software design (represented in the "Reuse of designs: an interdisciplinary cognitive approach" workshop -Visser, 1993- by only one paper, see Karakostas, 1993), other types of design (e.g., engineering, architectural) are well represented as application domains in the CBR research community.

As will be argued below, design -especially nonroutine design- is particularly suitable for case-based reasoning. An important aspect of expertise in -nonroutine- design is the construction, organisation and exploitation of a memory based on previous experience with particular design problems (a case memory). Psychological studies in the domain of analogical reasoning show that even experts have difficulty in accessing the case(s) relevant for solving a new problem (see Ross, 1990), so the system proposing cases to the user is one of the main assistance functions that should be provided by CBR-based tools. In addition, a critical component in such systems might assist designers in evaluating the relevance and suitability of cases. Moreover, a CBR system might help designers to (re)construct a case memory when new cases are added, and so assist these designers in their learning (by experience) (see Falzon & Visser, 1989) (see also, below, "Types of case use").

## 1. An interdisciplinary cognitive approach

Adopting an "interdisciplinary cognitive approach" to design activities not necessarily means that one seeks to implement, in computerised design systems, techniques which possess "cognitive validity" in an "emulative" sense, i.e. that the system emulates human designers' cognitive processing and structural characteristics.
In our view, it rather means that -based on the fundamental cognitive-ergonomics principle of "compatibility" between a system's functioning and its users' cognitive functioning- system specifications are to be based on those data concerning its user which are relevant in the context of system use. Generally, these are characteristics of mental representational processes and structures and of information processing.

The two may be related: a way to make computer systems compatible with their users' functioning may be to adopt, for the system's functioning, structures or processes "analogous" to those of its users. Such an "emulative" approach is not at all guaranteed, however, to result in the desired "system-user compatibility".

A few examples of relevant questions which might be examined in the context of an interdisciplinary approach to research on design reuse, are:

"How does the way in which cases are represented and organised in a case-based memory influence designers' activities (not only their retrieval of cases, but also their use of these cases in problem solving), and thus the resulting design?"

"How do different system retrieval modes influence designers' retrieval strategies and use of cases (and thus the resulting design)?" (see De Vries, 1993)

"How should one distribute the different tasks between a human designer and a system (e.g., what is difficult for human information processing systems, and which are their "good" aspects)?

Of course, the results of studies examining these questions do not have an immediate translation into specifications for CBR systems; the specifications they result in may lead to (technical) problems; a system based on these specifications - and thus based on "cognitive data"- may happen to offer no real assistance: the co-operation between user and system is still another point to be examined (see below).

## 2. Design & CBR

CBR systems have been proposed for several application domains: are there reasons to suppose that this form of reasoning is especially suitable for design?

### 2.1 Domains where CBR is applicable

Several papers presented at the workshop "Reuse of designs: an interdisciplinary cognitive approach" (Visser, 1993) start by a presentation of results of psychological research on human (design) problem solving activities, but they rarely show that - and/or how- these data have been taken into account, and nearly no author presents arguments (based on empirical data) to argue that, or why, design would be particularly suited to CBR.

The absence of a "strong domain model" is an often encountered criterion for a domain being suitable for CBR. The existence of such a model varies across design domains and the dimension routine - creative design (which is not necessarily confounded with the domain factor).

Referring to Spiro, Vispoel, Schmitz, Samarapungavan and Boerger (1987), De Vries states that "in ill-structured domains, crucial information tends to be uniquely contained in individual cases" (De Vries, 1993, p. 2).

It seems to us that design -an ill-structured problem solving task- and nonroutine design in particular, is especially suitable for case-based reasoning, because of the lack of design domain objects which could be instances of a same class -if the designer were able to proceed to the corresponding categorisation-, which could be covered by one and the same -not very general, abstract- model, or which could lead to the formulation of a rule possessing some generality.

**2.2 Empirical studies showing reuse in design activities**

Working in a KADS framework, Kruger (1993) has observed nine professional designers during their activity. She has noticed that they proceeded to reuse during various component design activities (or stages) ("select", "identify", "model" and "generate", p. d-2).

In three empirical studies on designers working in different domains (software, mechanical, and aerospace-structure design), Visser (1991, 1992b) has identified different aspects of design reuse.

• She has observed that designers both exploit existing design components in their problem solving under hand, and anticipate future reuse by creation of reusable structures.

• She has distinguished two different types of analogy use: at the action-execution level (actual problem solving), in order to solve the current design problem (the "classical" use of analogical reasoning), and at the action-management level (control), in order to make action-execution cognitively more economical (Visser, 1992c).

• However, not only analogy between a target problem and a case is exploited by designers in order to find a reusable design in memory, also other relationships are used (e.g., opposition and the relation between an action and its prerequisites) (Visser, 1991a).

• Cases at different abstraction levels are exploited: e.g., in software design, both at the code level and at the functional decomposition level (decomposition into different modules).

• Both "old" design solutions developed by the designers themselves and solutions designed by colleagues are referred to.

• Cases developed by the designers themselves can be solutions they have just designed (modules of the software under hand), or designs they have elaborated in a more or less remote past.

**3. <u>Distinctions between CBR and other data retrieval processes</u>**

CBR may be compared with other forms of data processing methods on different dimensions: e.g., the type of data used, the data retrieval mechanisms or the type of reasoning adopted (do CBR and AR, analogical reasoning, share this component? if yes, which aspect makes them different?), the applications for which they are most suitable.

### 3.1 Difference CBR - AR

Letia (1993) compares these two forms of reasoning with respect to the type of relationship between the target and the case and the different processing steps. Referring to Campbell and Wolstencroft (1990), Letia asserts that "structural similarity has been mostly used in AR, while organizational similarity is normally considered in CBR only, as a composition of semantic and pragmatic similarities" (p. b2). He adds, as a distinction between AR and CBR, that "CBR systems perform, instead of elaboration, mapping and inference, an adaptation, and the justification stage is emphasized as evaluation and repair. Therefore, CBR can be regarded as a special case -a weaker form- of AR." (Letia, p. b2)

Cunningham, Smyth, Finn and Cahill (1993) consider CBR and AR as two forms of "episode-based reasoning": CBR would concern intra-domain, whereas AR inter-domain remindings. The authors judge this demarcation "artificial". They consider that "the distinction is better characterised as a continuum of abstraction of remindings with CBR towards the concrete end and AR toward the more abstract end" (p. K-1).

They also notice that most work on AR handles with abstract, remote remindings based on abstract or thematic relationships, whereas in CBR remindings need a significant overlap of surface features. This results in a "relative shift in emphasis from base filtering (surface feature mapping) (in the case of CBR) to structural mapping (in the case of AR)" (p. K-2).
Still other differences between CBR and AR, displayed in their Figure 2 (see the paper), are the following:
• CBR: D-nets <-> AR: thematic hierarchy
• CBR: large role of base filtering <-> AR: small role of base filtering
• CBR: small role of mapping (simple shallow mappings) <-> AR: large role of mapping (more complex "structure mapping" techniques).

### 3.2 Difference CBR - reminding

Wharton and Lange (1993) contrast "computational models of general, non-expert reminding" (such as their REMIND) with "case-based reasoning models …, which usually simulate expert reasoning" (Wharton & Lange, p. D-1).

REMIND also differs from a CBR approach in that CBR generally uses "explicit, presumably conscious reasoning", whereas REMIND proceeds by "more subtle priming effects" (p. D-11).

### 4. <u>Knowledge in a system: only cases?</u>

Several systems combine case knowledge and models into "hybrid" systems. Bhatta and Goel (1993) adopt this integration for several reasons:
• both cases and models are available in domains such as physical devices design;
• "cases can quickly lead a problem solver into the neighborhood of a solution to the given problem and models can effectively guide it to adapt that solution to meet the requirements of the new problem. Thus an integrated system combines the powers of both case-based reasoning (i.e., efficiency) and model-based reasoning (i.e., efficacy)." (p. A-3)

The authors also integrate a model-based and a similarity-based approach to learning, their focus in this paper. Each one is used for a specific purpose. The model-based approach is used to learn the indexing features, and similarity-based learning to generalise over the learned features.

The FABEL project (Voss & Schmidt-Belz, 1993) is building an integrated knowledge and case administrative system with learning mechanisms for similarities.

Sycara and Navin-Chandra's (1993) system, CADET, uses cases and engineering domain laws and principles.

In their research on organic synthesis planning, Napoli and Lieber (1993) propose an integration of classification-based reasoning and case-based reasoning.

The rest of this paper only focuses on cases.

## 5. Cases

Cases, one of the central topics in CBR research, lead to different positions adopted by different authors and evoke many questions. They may be examined from different viewpoints; in this paper, several questions around different aspects are briefly alluded to: the use of cases, their format, their (de)composition, their type, and, of course, their organisation in a case memory.

### 5.1 Types of case use. Direct and indirect problem solving

In most CBR problem-solving systems, cases are used for, what we call, "direct" problem solving: a design problem is to be solved and a case is retrieved as a (part of a perhaps temporary version of a) solution to be adapted.

Cases may, however, also be used more "indirectly" in problem solving. An example of such "indirect" design problem solving is the use of cases to "stretch" the problem space (see De Vries, 1993), or -still more remote (in time) from the immediate problem-solving activity on a particular design project-, the use of a case base by (novice) designers to learn about solutions or other design-problem aspects, such as constraints. This issue is addressed by De Vries (1993) when she studies the kind of knowledge a designer can acquire from cases, in particular the degree to which previously used constraints can be learned from them. Inferring design constraints from old solutions is an example of using cases to "stretch" the initial problem space: "the set of constraints that has been proven useful for developing qualitatively good solutions should be used in future design" (De Vries, p. B-2).

The distinction is, of course, only a relative one: should solution evaluation be considered "direct" or "indirect" problem solving? Nakatani, Tsukiyama and Fukuda's (1993) prototype system SUPPORT provides designers with cases to assist them in both solution development and solution evaluation: "at the end of each [solution development] step, constraint violations are checked. If some constraints are violated, the system and the user interactively try to recover them based on similar violations in the past" (p. H-1).

### 5.2 Components of a case

Different authors decompose -or compose- their cases differently.

In FABEL (Voss & Schmidt-Belz, 1993) a case consists -for the moment, according to the "compromise case format" adopted- of the following "components":
• a problem situation (initial state)
• a problem specification
• the solution situation (goal)
• the way of elaboration (sequence of operations)
• unstructured annotations.

Voss and Schmidt-Belz work on architectural design. They wonder how to take into account the geometrical relations between cases: their "expert" uses various global-pattern based descriptions ("string of pearls", "comb", "herringbone").

CADET (Sycara & Navin-Chandra, 1993) manages composite cases, aggregating the different retrieved components.

In Bhatta and Goel's (1993) IDeAL (Integrated Design by Analogy and Learning), a design case specifies the functions delivered and the structure, and has a pointer to the causal behaviours (the "structure-behaviour-function" model).

### 5.2.1 Does one need the "context" of a case?

Voss and Schmidt-Belz (1993) observe that designers, (sometimes) in order to be able to use a case, seem to look for, or need, its context. They wonder if this context has to be stored and/or represented explicitly together with the case (as part of the case), or if cases may be stored and/or represented in such a way that designers can retrieve the case context -if they want to.

### 5.2.2 "How to cut a complex design into cases?"

To a psychologist, this question -asked by Voss and Schmidt-Belz (1993)- seems a "computer-science" question -if the answer does not make any reference to human designers. Psychologists would examine this problem from two different viewpoints:

• If designers proceed to such decompositions, how do they <u>actually</u> decompose their design knowledge, their design problems and solutions? This question would be examined through observations (in controlled, experimental, or more exploratory field studies).

• What do designers want -or (think they) need- to store as a "case", a "composite case", or "case component"? This question would be examined through interview studies.
(see Visser, 1992a, for the differences between these two approaches -and other ones)

### 5.3 Types of case

The main question with respect to this point concerns the specific-general nature of cases, but other types are also distinguished.

### 5.3.1 Particular or general

In introductory (sections of) papers on CBR or reuse, the "specific" character of objects reused (cases) is often token as characterising these forms of reasoning. The trend in the domain of CBR to propose "mixed" or "combined" systems also presupposes that "cases" are specific. Many researchers, however, do not seem to adopt strictly this position.

According to Kolodner (1989), cases may be "particular", or "generalized or composite" ones ("used over and over with only the variables changed each time") (p. 156). Two other authors who stood -as fathers- at the cradle of the CBR approach, Riesbeck and Schank (1989, p. 11) have proposed three major types of case: rules ("ossified cases"), cases ("paradigmatic" cases and their "affiliated" cases), and stories. If the differences between these cases -according to the authors, the way they are encoded in memory- are not very clear, their proposals may have a heuristic function.

### 5.3.2 Other distinctions

One may consider that there are still other "types" of case which possibly coexist in a case base.

Voss and Schmidt-Belz (1993), e.g., distinguish "solution cases", "strategy cases", "similarity concepts" and "snapshots".

Snapshots are "holistic cases", "special restricted views on a much more complex situation, suitable for solving or communicating about a particular problem". A snapshot expresses different -interconnected- problems and solutions, and different snapshots might be linked by elaboration relations.

De Vries (1993) asks if cases are always "old" solutions, or if "old" constraints can also constitute cases. Kruger (1993) notes that reuse is not limited to reusing "old" particular design solutions, it also involves reuse of "knowledge about the situation the object performs in, including knowledge about the behavior of people in this situation" (p. d-2) (see De Vries, 1993, who studies how this type of knowledge puts constraints on a design).

<u>Problem - Solution</u>. For a cognitive psychologist, "solution" is a very general and relative notion. In Visser (1991), we have analysed the couple "problem" - "solution" from four viewpoints:

• "Problem" or "solution"? A same entity may be conceived as a "solution" or as a "problem," depending on whether it is an output or an input to a problem-solving action.

• A solution "unit"? A knowledge unit may constitute a "solution" to a "problem" without being coded as such in memory; still, people do have units in memory integrating knowledge about problems and their solutions: these units may concern classes ("problem/solution schemata"), or specific problems and their solution(s) ("problem/solution associations") (cases).

• The scope of a solution: when a problem/solution is activated, what is activated? It certainly is neither only its "label", nor everything a person "knows about" it.

• Levels and viewpoints of solution representation: problem solving involves processing different types of representation at different levels and from different viewpoints. A design is developed (a) in greater and greater detail: first defined in terms of a global goal to be achieved, it is progressively characterised to become sufficiently precise as to specify the realisation of the target object; (b) from different viewpoints which are all to be taken into account: e.g., functional, structural, topological, behavioural, material / physical, financial.

### 5.4 Organisation of case memory

The organisation of a case base is of course important from an AI architectural viewpoint, but also from a user's viewpoint. Indeed, according to the links existing - or dynamically established- between cases, a designer may access several "related" cases: especially for "indirect" case use (see above), this may be useful.

Nakatani et al. (1993) divide their casebase "into several sub-casebases according to their designing years".

The concept of "dynamic memory" (Schank, 1982) refers to "a memory system which changes and reorganises itself with new experience". Several authors refer to this approach to memory.

Bhatta and Goel (1993) propose a hierarchical organisation of case memory based on multiple generalisation-specialisation hierarchies (in reference to Kolodner, 1980) (in which "multiple" refers to the multiple dimensions pertaining to the constituents of device function and structure).

Cunningham et al. (1993) work with D-nets (discrimination networks). They criticise them for pragmatic -but not principle- reasons. "In straightforward CBR systems, discrimination networks … are an effective method for organising the case memory to support base filtering" (p. K-1). However, when "remindings are more abstract, or [when] the system may need to support remindings at different levels of abstraction, it becomes more difficult to organise the case-base as a D-Net. The argument sometimes presented is that this dependence on indices will preclude remote remindings" (p. K-1).

"Indexed memory will present difficulties in supporting remote or abstract remindings". In this case, "retrieval is only possible when cases are stored without indexing. This requires that memory is content-addressable in that all information about stored cases is matched with the target case and the 'best' match is returned. It is assumed that the exhaustive search implied in this approach is made feasible by parallel hardware." (p. K-3)
In a note, the authors mention that, "besides this 'anti-indexing' research, there are also 'anti-discrimination networks' arguments within the indexing school. The alternatives to retrieval based on D-nets in indexing are parallel algorithms of spreading activation" (ibid.). For routine design, the authors nevertheless stick to their organisation based on D-nets.

In W. Visser (Ed.), *Proceedings of the IJCAI Thirteenth International Joint Conference on Artificial Intelligence Workshop "Reuse of designs: An interdisciplinary cognitive approach", Chambéry, France, August 29, 1993* (pp. 1-14). Rocquencourt, France: Institut National de Recherche en Informatique et en Automatique.In nonroutine design, (still more) remote, abstract remindings are, however, generally required. The authors reject two possible solutions in the domain of "current automatic base filtering techniques": "multi-level indices" and the use of disjoint nets for each level in the index hierarchy. Their solution in these domains is "case retrieval assistance" rather than "automatic support". "Index transformation is a suitable point where man-machine interaction can usefully occur" (p. K-9). It "allows specifications to be interpreted from different perspectives. The user can provide the context sensitivity that can highlight certain features, de-emphasize others, and expose hidden or implicit goals." While the authors judge this task "difficult for a computer", they think it "particularly easy for a human" (p. K-9) (compare this position with the approach taken by Sycara and Navin-Chandra, 1993, using automatic index transformation).

Cunningham et al. (1993) examine four modes of index transformation: abstraction, specialisation, elaboration and condensation. Visser (1991) has observed designers using a combination of abstraction and specialisation (an approach she entitled "going up and then down again in the solution abstraction hierarchy") and she described it as involving "a solution development in at least two steps: elaboration, first of an alternative solution at a higher abstraction level, and then of a more specific solution to the problem constituted by the abstract solution" (Visser, 1991, p. 694).

"Elaboration and condensation" also refer to design activities considered important in recent empirical studies on design (see Visser & Hoc, 1990): as noticed by Cunningham et al. (1993), specifications are often vague, and so they fail to bring about the appropriate remindings, or bring about excessive remindings.

<u>Availability ≠ Accessibility</u>. Labelling (indexing) is also such a crucial aspect of case-based memories because, according to the possibility to establish relevant connections between the designer's knowledge representation and the information in the base, it conditions the <u>accessibility</u> of the information which is <u>available</u> in the case memory (see De Vries, 1993, referring to Oxman, 1992) (see also Smaïl, 1993, when she distinguishes "precision" and "recall").

De Vries (1993) (in reference to Eylon & Reif, 1984) notes that there are "studies on learning [which] provide arguments for a hierarchical organization of information"; other studies, however, e.g. "Spiro et al. (1987), … argue that knowledge representations suitable for transfer … should be flexible, assembled in order to fit a context, interconnected, complex and stick to specific details of particular cases. Nonlinear and multidimensional texts would permit criss-crossing from case to case in many directions with many thematic dimensions. According to Spiro et al. linear presentation should be more suitable for mnemonic recall, whereas a network structure would be more suitable for transfer and application of the information." (p. B-7)

Access through keywords would not be appropriate: e.g., it is a static method and it requires that the user knows the keywords (see De Vries, 1993, referring to Fischer, McCall & Morch, 1989).

The results obtained in her case-base exploration experience led De Vries (1993) to conclude that "exploitation of a case memory for idea generation and for stretching the problem space for problem solving … can best be supported by some kind of network structure that permits free navigation from case to case" (p. B-11). She notices also that "other tasks, such as searching for a specific case, may demand some kind of indexing" (ibid.).

<u>Individual vs. collective case memory</u>. De Vries (1993) addresses the question related to the influence of the organisation of cases in external memory and the type of retrieval activity on the designer's problem space for problem solving. "External case memory, whether electronic or paper, containing known solutions to architectural design problems constitutes the collective memory within the domain." (p. B-2)

## 6. <u>Case retrieval</u>

In CBR systems, most retrieval proceeds by indexes (labels), but these may be of various types; CBR systems use retrieval processes and strategies, and so does a human problem solver, but • both use not necessarily the same, and • if a CBR system is an assistance system, the human problem solver will have to use specific strategies to use the system's retrieval functions.

### 6.1 Indexes

Cunningham et al. (1993) discuss four types of index of "increasing semantic separation" in an "index abstraction hierarchy" with different abstraction levels:
• "surface indices reflect observable features in a case";
• "behavioural indices embody emergent behaviour of the case stemming from the interaction of the components within the case …. It is at this level that matching between the electrical and mechanical systems could be achieved";
• "causal indices capture explanatory knowledge about how the behaviour comes about as a consequence of the interaction of the sub-entities within the case";
• "teleological indices detail the purpose or goal of the case" (pp. K5-6).

In FABEL (Voss & Schmidt-Belz, 1993), different kinds of indexes are used : surface descriptors (symbolically represented features: attribute-value pairs), pixel maps resulting from scanned 2D representations, design process operators and/or spatial-dependency graphs.

In Bhatta and Goel's (1993) IDeAL, a design case is indexed by task specifications (by structural and functional indices). The authors focus on index learning -done automatically- which is necessary because of the dynamic character of case knowledge resulting from problem-solving experiences.

<u>Use of qualitative knowledge for indexing</u>. Sycara and Navin-Chandra's (1993) CADET does not use a "pre-defined set of indices to access cases in memory" (J-6). It uses a "technique that can recognize underlying behaviorial similarities" (J-6), combining (a) "behavior-preserving transformation techniques that convert an abstract description of the desired behavior of the device into a description that facilitates a similarity recognition and subsequent retrieval of relevant components" and (b) "matching and extracting of relevant snippets of designs in the case-base" (J-6).
Several models add or elaborate qualitative knowledge to facilitate the retrieval of relevant cases. For example, CADET's matching module elaborates influences such as: *rotation <---+-- pressure*, and uses them as an index.
The authors judge that the transformation technique used in CADET "is applicable to any domain in which behavior can be modeled as a graph of influences" (J-8).

### 6.2 Relations retrieval is based on

Different systems use different relations for their retrieval.

"Similarity between target and source" is, according to Cunningham et al. (1993), the relation used by both CBR and AR for their retrieval of episodes from memory.

Nakatani et al. (1993) use "conceptual similarity" relations.

Kruger (1993) observes that human designers' access to a "case" seems to be based not only on the case's "primary" function (i.e. the actual function of the corresponding object), but also on "secondary", possible functions (p. d-3) (cf. approaches in terms of different "viewpoints" on a design object -e.g. Visser, 1992b- or in terms of the "salient" character of features according to the task in which it is used).

Wharton and Lange's (1993) "REMIND can explicitly create inferences that are only implied in the text representation of the cue" (p. D-1). This allows the model to do abstract, "cross-contextual" reminding, i.e. retrieval of episodes that do not share similarity [with the 'memory probes'] [i.e. little surface featural overlap in terms of situations or characters] except at the level of higher order relations such as a similar theme" (ibid.). In the psychological experiment presented in their paper, the authors have observed that their "subjects" indeed use this type of reminding.

### 6.3 Retrieval and selection

In general, one is not interested to retrieve "too many" cases. Most systems therefore proceed in two steps, retrieval followed by selection, or selection followed by retrieval. Kolodner (1989) distinguishes three major ways in which researchers in the case-based reasoning community are addressing this selection problem:
1. use indexes as selectors: try to determine how to best choose indexes so that only the best cases will be retrieved from memory;
2. filter the problem description before probing memory;
3. filter after retrieval.

Kolodner adopts the third way, but whereas "the methods [used by others] all tend to be special purpose", she proposes a general approach. Her two retrieval steps are:
• retrieval of partial matches;
• selection of the "best" matches;
The first retrieval step is based on similarity; among the various retrieved cases, a selection is established using preference heuristics, two of which address usefulness.

So Kolodner (1989) proceeds by "selection of the best" using "usefulness rather than similarity". She judges that "it only makes sense to focus on relative goodness of correspondence [based on similarity] after we know that the reasoner's goals are being attended to" (p. 161). "Emphasis in choosing best cases is on <u>usefulness</u> rather than similarity" (p. 155). The adoption of the "usefulness" criterion means that one takes into account the reasoner's goals. Rather than working with indexes used as "restrictors", she proposes feature sets used as "selectors". Her system's memory "prefers cases that match on salient features", but "if no cases matching on salient features can be found in memory", "it also allows recall based on features that were not singled out at memory update time" (p. 161).

If retrieval leads to too many relevant cases, CADET (Sycara & Navin-Chandra, 1993) selects the "best" case (a term not used by the authors) based on "criteria that prefer alternatives that exhibit design simplicity, ease of synthesis, and low cost" (p. J-11).

Smaïl (1993) distinguishes the proportion of retrieved items actually relevant ("precision") from the proportion of relevant items actually retrieved ("recall").

Simoudis and Miller (1990) introduce "selection by 'validation'": a "'simple' retrieval" stage (based on similarity of surface features) is followed by a domain-specific "validation": one applies to the target problem the tests of the retrieved cases for which one knows the values that must be met for the case to be valid; the results are compared with these values. Using this approach, selectivity augments from 11% (without validation, i.e. using only "simple retrieval") to 1.5% to 3% (with validation, i.e. using "validated retrieval").

**6.4 Retrieval processes**

How can one be sure to retrieve (only) the "best" cases? Voss and Schmidt-Belz (1993) think that probably different, various retrieval techniques are necessary, based on different types of representation feature used to index.
They wonder if one should have various different static underlying representations (and if yes, how many?), or if each retrieval mechanism should generate dynamically its own representation (cf. the trade-off between redundancy and computation time).

De Vries (1993) is interested by human retrieval strategies. "Users may or may not have a definite target in mind when consulting a case memory. This constitutes the main ground for distinguishing between different retrieval strategies" or types of "system use" (p. B-7). She distinguishes "search" and "browsing" as two different retrieval strategies: when a person "searches", "some fact is the target of system use", while "browsing" means that "no such definite target is present" (p. B-7).

The results of her experiences are that, compared with searching, browsing leads to more stretching of the initial problem space.

Two remarks may be formulated concerning these results:
1. the difference is statistically significant, but it is not large (the mean number of new constraint topics recalled after the experimental task is 6.9, compared with M = 6.1 before);
but more importantly:
2. the experimental task may be thought to induce browsing because the target of system use has been defined, not by the subjects themselves, but by an external source (the experimenter).

## 7. Development of CBR systems

The main AI questions for CBR assistance systems might also be organised at the two levels distinguished by Newell (1982) (see Trousse & Visser, 1993).

### 7.1 CBR systems: assistance or autonomous systems

Bhatta and Goel's (1993) IDeAL system is an autonomous design system for the design of physical devices. Other authors propose assistance systems. Nakatani et al.'s (1993) prototype system SUPPORT is an interactive assistance system providing designers with support in several conceptual-design stages ("specification analysis", "functional" design, "device parts selection" and "device parts layout").

The architect studied by Voss and Schmidt-Belz (1993) has made a move from autonomous to intelligent assistance systems. Experience with previous knowledge-based system prototypes led him to "abandon the idea of automating architectural design and to turn to case-based reasoning combined with modest model-based or heuristic support" (p. C-2).

Sycara and Navin-Chandra (1993) propose that, because "CADET can generate a wide variety of behaviorally equivalent alternative designs for a set of design specifications, it can be used as a designer's brainstorming assistant" (p. J-1), "not his replacement" (p. J-5). "Verification and debugging is currently left to the human designer. [CADET]'s usefulness lies in its ability to access a large database of prior designs and find relevant components and alternative configurations." (p. J-5)

For Cunningham et al. (1993), nonroutine design needs "interactive case-based design assistants", not autonomous, automatic systems.

Letia (1993) believes that "a knowledge based design assistant should provide CBR/AR related services like:
• show the distribution of cases in the problem space,
• retrieve the most similar case,
• retrieve the most similar and most different case,
• delete a case,
• ask permission to learn a new case,
• generate features different from those of current case,
• show the current indexing function,
• modify the current indexing function,
• provide adaptation rules modifiable by the designer.
It would appear that a suitable organization might require a multiple indexing scheme.
Adaptation, evaluation and repair are better left, particularly in non-standard situations, under the control of the designer." (Letia, p. b3)

In order to guarantee a certain cognitive validity of the CBR assistant, one needs to take into account the user-machine co-operation from the knowledge acquisition phase until the design of the system. Some examples of questions which may guide this development are:

• Which co-operation models could one imagine for systems reasoning with cases, and how are these linked to the routine-creative dimension of design?

• What types of knowledge should be provided to a CBR assistant in a specific design task?

• Which are the consequences on the architecture of a CBR system, and on the flexibility the system offers the user?

**7.2 Case knowledge acquisition**

Two major problems are the following.

• How should cases, the way they are used, and the way they might be used most profitably, be "acquired"?

• Can case-oriented acquisition be integrated into existing knowledge acquisition methodologies (such as KADS); if yes: how? if not: how should knowledge acquisition methodologies be modified?

The FABEL project -presented by Voss and Schmidt-Belz (1993)- tackles this point. One of its main purposes is indeed "to explore how cases can be used … during knowledge engineering" (p. C-1) Adopting the KADS framework, the authors aim at a "case-oriented knowledge acquisition methodology" in which "KADS marries cases" (p. C-11).

A methodological problem with respect to identifying the actual use, or the type of use, designers make of cases in their activity is the following: it may occur that designers who are asked to verbalise their thoughts simultaneously with their problem-solving activity, "mention" a problem and/or solution without the experimenter being able to infer -from the way the case is verbalised or from contextual elements- if the case is being used -or has been used- directly or indirectly, in the solution-development activity, or if it is only referred to in order to explain or describe a solution developed by another type of reasoning (see Kruger, 1993). Falzon and Visser (1989) have identified this problem as a general one in knowledge acquisition, especially when the traditional -and in AI preferred- interview methods are used: designers may present their knowledge (structures and processes) in terms adapted to the knowledge engineers. The authors concluded that "naïve physics may be an adapted answer for naïve [knowledge engineers]!"

**7.3 Empirical evaluation of a CBR system**

This point is seldom discussed. Nakatani et al. (1993) report concerning the evaluation of their prototype system that its use reduces a conceptual-design task which manually took more than 10 hours, to four or five hours.

Other aspects should be considered in order to evaluate the assistance offered by a system and to improve its (future) cognitive validity. A point to be addressed in particular is: how, and what, can one learn from the interaction between the user and the system? In this context, Smaïl (1993) proposes an interactive and iterative retrieval process, and uses learning mechanisms for relevance feedback.

## 8. Conclusion

The "Reuse of designs: an interdisciplinary cognitive approach" workshop (Visser, 1993) was supposed to focus on two aspects of reuse:

• its interest for nonroutine design tasks in which a human designer and an intelligent artificial system collaborate, and

• the need to consider data on cognitive activities involved in these tasks so that the assistance system can be made compatible with the human information processing system.

The papers presented at the workshop address various important points, but those presenting CBR systems generally focus on the system, its processes and structures, without paying much attention to its users -even if the systems are supposed to be assistance systems.

Few papers have adopted an interdisciplinary and/or cognitive approach. The "collaboration" between a human information system and an AI system is generally based on pragmatic reasons: autonomous, automatic systems do not manage alone in complex tasks, and this problem is handled in a pragmatic, empirical way -even if this stand is never made explicit.

In W. Visser (Ed.), *Proceedings of the IJCAI Thirteenth International Joint Conference on Artificial Intelligence Workshop "Reuse of designs: An interdisciplinary cognitive approach", Chambéry, France, August 29, 1993* (pp. 1-14). Rocquencourt, France: Institut National de Recherche en Informatique et en Automatique.

Sycara, K. & Navin-Chandra, D. (1993). Case representation and indexing for innovative design reuse. In W. Visser (Ed.), Proceedings of the Workshop of the Thirteenth International Joint Conference on Artificial Intelligence "Reuse of designs: an interdisciplinary cognitive approach", Chambéry (France), August 29, 1993.

Trousse, B. & Visser, W. (1993). Use of case-based reasoning techniques for intelligent computer-aided design systems. Proceedings of the IEEE/SMC'93 Conference - System, Man and Cybernetics - Systems Engineering in the Service of Humans, Le Touquet, France, October 17-20, 1993.

Visser, W. (1991). Evocation and elaboration of solutions: Different types of problem-solving actions. An empirical study on the design of an aerospace artifact. In T. Kohonen & F. Fogelman-Soulié (Eds.), COGNITIVA 90. At the crossroads of Artificial Intelligence, Cognitive science, and Neuroscience. Proceedings of the third COGNITIVA symposium. Amsterdam: Elsevier.

Visser, W. (1992a). Data collection methods in cognitive psychology: A presentation through the example of design activity studies. Actas 1991. Programa LINGUA (Colecção "Temas Educacionais"). Lisboa: Universidade Aberta, Centro de Estudos de Ensino a Distância.

Visser, W. (1992b). Designers' activities examined at three levels: organization, strategies & problem-solving. Knowledge-Based Systems, 5 (1), 92-104.

Visser, W. (1992c). Use of analogical relationships between design problem-solution representations: Exploitation at the action-execution and action-management levels of the activity. Studia Psychologica, 34 (4-5), 351-357. (abstract in International Journal of Psychology, 1992, 27, 156)

Visser, W., & Hoc, J.M. (1990). Expert software design strategies. In J.M. Hoc, T. Green, R. Samurçay & D. Gilmore (Eds.), Psychology of programming. London: Academic Press.

Visser, W. (Ed.). (1993). Proceedings of the Workshop of the Thirteenth International Joint Conference on Artificial Intelligence "Reuse of designs: an interdisciplinary cognitive approach", Chambéry (France), August 29, 1993.

Voss, A., & Schmidt-Belz, B. (1993). Case-oriented knowledge acquisition for architectural design. In W. Visser (Ed.), Proceedings of the Workshop of the Thirteenth International Joint Conference on Artificial Intelligence "Reuse of designs: an interdisciplinary cognitive approach", Chambéry (France), August 29, 1993.

Wharton, C. M., & Lange, T. E. (1993). Case-based retrieval and priming: Empirical evidence for integrated models. In W. Visser (Ed.), Proceedings of the Workshop of the Thirteenth International Joint Conference on Artificial Intelligence "Reuse of designs: an interdisciplinary cognitive approach", Chambéry (France), August 29, 1993.

18